\documentclass[Review,sageh,times,fleqn]{sagej}

\usepackage{moreverb,url}

\usepackage{subfigure}

\usepackage[table]{xcolor}

\usepackage[colorlinks,bookmarksopen,bookmarksnumbered,citecolor=red,urlcolor=red]{hyperref}

\newcommand\BibTeX{{\rmfamily B\kern-.05em \textsc{i\kern-.025em b}\kern-.08em
T\kern-.1667em\lower.7ex\hbox{E}\kern-.125emX}}

\def\volumeyear{2019}

\begin{document}


\title{Terrain estimation via vehicle vibration measurement and Cubature Kalman Filtering}

\author{Giulio Reina\affilnum{1, 2} and Antonio Leanza\affilnum{2} and Arcangelo Messina\affilnum{2}}

\affiliation{\affilnum{1} Politecnico di Bari, Department of Mechanics, Mathematics \& Management, Italy \\ \affilnum{2} Università del Salento, Department of Engineering for Innovation, Italy}

\corrauth{Giulio Reina, Politecnico di Bari, Department of Mechanics, Mathematics \& Management, Via Orabona 4, 70125 Bari, Italy.}

\email{giulio.reina@unisalento.it}

\begin{abstract}
The extent of vibrations experienced by a vehicle driving over natural terrain defines its ride quality. Generally, surface irregularities, ranging from single discontinuities to random variations of the elevation profile, act as a major source of excitation that induces vibrations in the vehicle body through the tire-soil interaction and suspension system. Therefore, the ride response of off-road-vehicles is tightly connected with the ground properties.\\
The objective of this research is to develop a model-based observer that estimates automatically terrain parameters using available onboard sensors. Two acceleration signals, one coming from the vehicle body and one from the wheel suspension are fed into a dynamic vehicle model that takes into account tire/terrain interaction to estimate ground properties. To solve the resulting nonlinear simultaneous state and parameter estimation problem, the Cubature Kalman filter is used that is shown to outperform the standard Extended Kalman filter in terms of accuracy and stability. An extensive set of simulation tests is presented to assess the performance of the proposed estimator under various surface roughness and deformability conditions.\\
Results show the potential of the proposed observer to estimate automatically terrain properties during operations that could be implemented onboard of a general family of intelligent vehicles, ranging from off-road high-speed passenger cars to lightweight and low-speed planetary rovers.
\end{abstract}


\maketitle
\runninghead{Terrain estimation via Cubature Kalman filtering}

\section{Introduction}

Terrain properties have a significant impact on traction, ride and handling quality, and energy consumption of off-road heavy/light vehicles. Adhesion coefficient (for on-road vehicles) and soil cohesion and internal friction angle (for off-road vehicles) control wheel traction properties, with implications for longitudinal and lateral wheel slip \citep{REI1}. Additionally, terrain deformability modulates the vibrations induced in the body of heavy vehicles by terrain irregularity \citep{PAR} affecting the ride quality and the vertical forces acting on a tire, which in turn impact on vehicle handling characteristics \citep{WON}. Power demand is also tightly connected with terrain type. This aspect is especially important to increase the autonomy of self-driving and battery-powered robots, like planetary exploration rovers \citep{SPO}.\\ Therefore, the ability to automatically estimate the properties of the supporting surface traversed by a vehicle would be an enabling technology with many useful applications. For example, automated strategies may be implemented to ensure safe driving. Knowledge of terrain properties could be also used for adaptive regulation of traction control systems, assisting braking systems, and active or semi-active suspensions. Driving energy estimation algorithms could be also adopted to develop strategies for energy-efficient path planning.\\
This paper presents a non-linear model-based observer for combined estimation of terrain properties and ride motion states. The algorithm operates on noisy sensory data obtained from a body-mounted and a suspension-mounted accelerometer that are standard equipment for modern automobiles and mobile robots. The algorithm relies on the observation that distinct terrain types possess different elevation profiles and bulk deformability properties, which give rise to unique, identifiable acceleration signatures during the interaction with a rolling tire. The authors have previously proposed a general quarter-car model suitable for off-road applications that takes into account tire-soil interaction. The model can handle the general case of a compliant wheel that rolls on compliant ground and it allows ride and road holding performance to be evaluated \citep{REI2}. In this research, the general quarter-car dynamic model forms the basis of the combined state and parameters estimation problem that is solved resorting to the novel Cubature Kalman filtering.\\

\section{Related Research}
The Kalman filter is the de-facto solution to estimation problems. It provides the optimal solution in terms of minimization of the mean square estimation error under the assumptions of linear model and Gaussian noise with many applications in the automotive \citep{DOU} and mobile robotics field \citep{REI16}. However, it is difficult to obtain a closed-form solution for the optimal nonlinear filter and some approximations are required. The Extended Kalman filter (EKF) can handle nonlinear state modeling \citep{BIS}. The EKF requires calculation of Jacobian matrices, which are the first order approximation of nonlinear process and measurement models. Jacobian calculation can be a non-trivial task, especially for highly non-linear systems and it increases the overall computational burden. Thus, the accuracy of EKF depends on the truncation errors introduced by the linear approximation. The higher order moments neglected by the EKF approach may lead to filter divergence, narrowing the applicability in practical nonlinear problems \citep{JUL1}.\\
These issues can be addressed by resorting to Unscented Kalman Filtering (UKF) that computes the carefully selected ``sigma-points" through the nonlinear transform in order to obtain the mean and covariance with higher accuracy than EKF \citep{JUL}. It does not need to linearize the system and measurement equations as required by the EKF.
Recently, \citep{ARA1} proposed the Cubature Kalman Filter (CKF) that was shown to improve the performance over UKF, as well as others nonlinear Kalman filters (e.g. Central-Difference Kalman Filter (CDKF)). Since nonlinear filtering can be reduced to a problem of how to compute integral, the CKF introduces a third-degree spherical-radial cubature rule to achieve the cubature points which are used to approximate the multidimensional integral. For nonlinear system with additive Gaussian noise, the CKF can preserve the first and second moments more accurately than the UKF with similar or lower computational complexity and run time. Moreover, even if both the UKF and the CKF use a weighted set of symmetric points, the CKF only requires $2n$ cubature points with equal weight, whereas the UKF needs $2n+1$ sigma points with a central point that is weighted more than the others.\\
Examples of non-linear estimation problems in the general field of driving automation can be found in \citep{RAY} where an EKF observer was proposed to estimate side-slip angle using wheel forces as input. An EKF framework was adopted as well as in \citep{BES} and \citep{BLA} to identify handling dynamic states, whereas a dual EKF was introduced in \citep{WENZ} for vehicle state and parameter estimation. The use of UKF was proved for vehicle state estimation in \citep{ANT}.\\CKF has been proposed and used in many applications, including positioning \citep{PES} and attitude estimation \citep{LI}.\\
The novel contribution of this research is manifold. First, a single parameter referred to as the equivalent soil stiffness is introduced to define collectively the terrain properties. It is estimated by observing the vehicle ride response using standard sensors and a dynamic model that takes into account the mechanisms underlying wheel/terrain interaction. Secondly, an estimator grounded in the cubature integration rule is proposed that ensures higher accuracy and stability than standard EKF. The observer is based on a general vertical dynamic model of an off-road vehicle, which is another novel contribution to the Literature. Thirdly, a solution to the joint vehicle state and parameter estimation problem using CKF is proposed by augmenting the state vector of the filter with the equivalent soil stiffness.\\
Other approaches have attempted to estimate terrain properties by collecting various sets of features through the onboard sensor suite that are used to train a ground classifier that uses a machine learning algorithm including Support Vector Machine (SVM) \citep{IAGN1, REI6}, Bayesian network \citep{GAL}, and Deep Learning \citep{RAM}. Here, a model-based approach is used where a general vertical dynamic model is adopted. Following this rationale, a better understanding of the physical mechanisms underlying vehicle-terrain interaction can be achieved. In addition, the influence of the many parameters entering the problem can be properly assessed.\\

%
%

\section{Vehicle-terrain interaction modeling}

\begin{figure}[hbtp]
\centering
\includegraphics[scale=0.1]{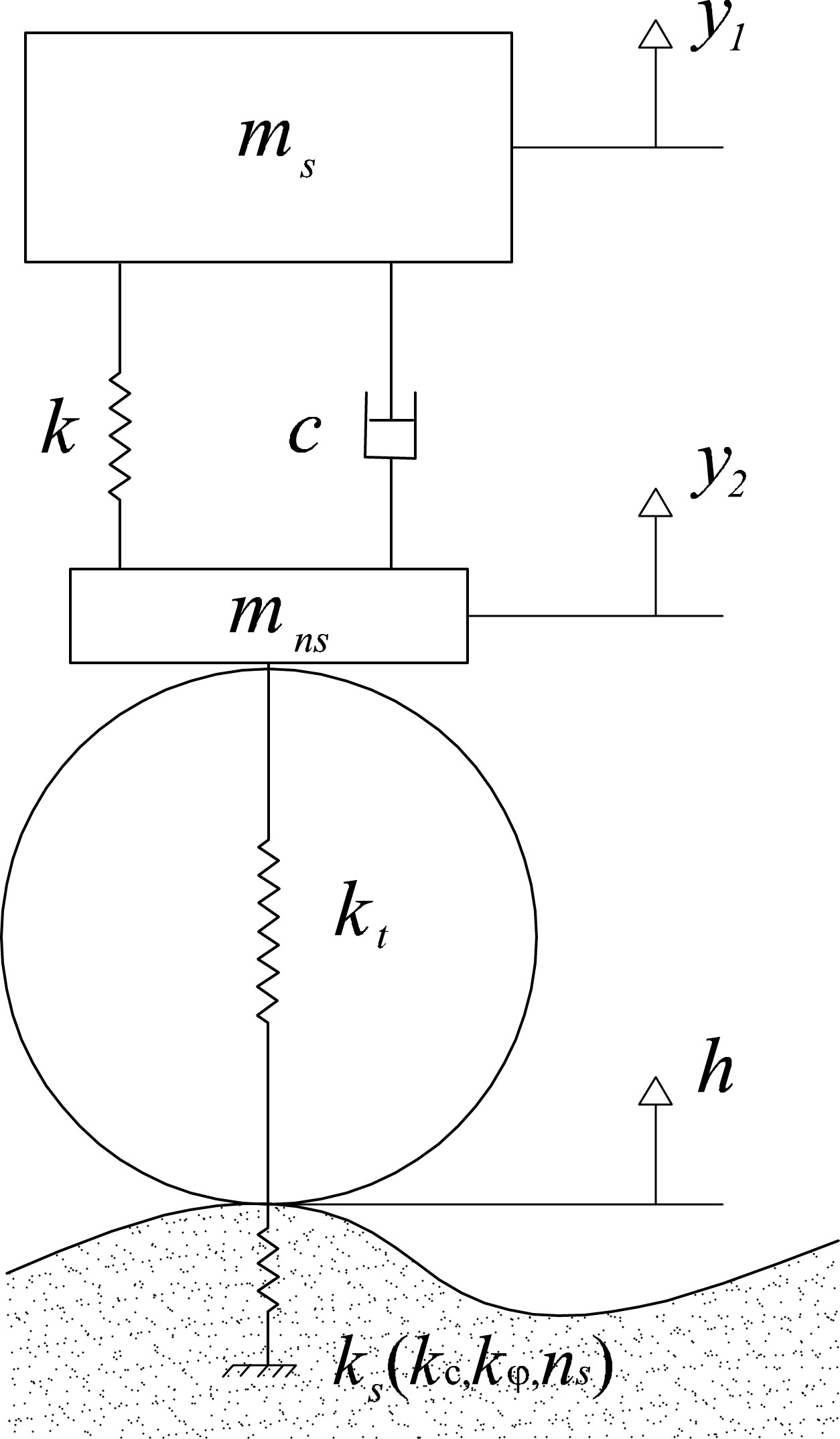}
\caption{General 2 DOF QC that includes the effects of terrain deformability}
\label{AugmentedQC}
\end{figure}

Vehicle ride response is usually studied by referring to a Quarter Car (QC) model \citep{REI}. However, the standard QC model neglects the deformability of the supporting surface and its applicability is limited to the traverse of hard terrain. In previous research by the authors, a general QC was proposed that included the effects of terrain deformability by modelling tire-ground interaction drawing on classical Terramechanics theory. Here, the general QC model is briefly recalled, whereas the interested reader is referred to \cite{REI2} for more details. With reference to Fig. \ref{AugmentedQC}, the vertical dynamic response can be studied in terms of the two degrees of freedom (DOF), namely, the vertical displacement of the vehicle sprung mass, $m_s$, and of the unsprung mass, $m_{ns}$, that are indicated, respectively with $y_1$ and $y_2$. Elastic and damping properties of the suspension are expressed, respectively, by $k$ and $c$, whereas the tire stiffness is indicated by $k_t$. This parameter set is fixed by design and it is not subject to temporal changes.\\ Soil deformability is accounted for by introducing an equivalent stiffness, $k_s$, that works in series with the tire. Then, the corresponding dynamic component of the force that the ground applies to the tire can be expressed as
\begin{equation}
F_g^d=k_{s}(k_\phi,k_c,n_s, W, D, b)z_r
\label{terra1}
\end{equation}
$z_r$ being the dynamic sinkage, that is the relative displacement with respect to the static sinkage. Given the vertical load $W$, and the geometry of the wheels (diameter $D$ and width $b$), the equivalent terrain stiffness $k_s$ results in a non-linear function of the Bekker-Reece terrain properties: the modulus of frictional soil stiffness $k_\phi$, the modulus of cohesive soil stiffness $k_c$, and the modulus index of soil deformation $n_s$. The value of these parameters can be found for common terrains in the Literature, for example in \cite{WON}.

Equation (\ref{terra1}) can be linearized in the vicinity of the static equilibrium point via Taylor series expansions without losing much accuracy (i.e., approximation errors less than $2\%$ for $40\%$ change in the dynamic ground force \citep{REI2}). Therefore, the linearized equivalent stiffness approximates quite well the relationship between ground force and relative sinkage. In addition, the introduction of $k_s$ allows the terrain properties to be defined collectively by a single convenient parameter. By estimating $k_s$, it is possible to infer the type of terrain that the vehicle is traversing. As an example, the value of $k_s$ for typical terrains encountered by a cross-country vehicle is collected in Table \ref{terrains}.

\begin{table}[hbtp]
\centering
\begin{tabular}{|l|l|}
\hline
\textbf{Terrain type}   & \textbf{$k_s \hspace*{0.1cm} (\frac{kN}{m})$} \\ \hline
Upland sandy loam       & 218.1            \\ \hline
LETE sand               & 2283.0           \\ \hline
Rubicon sandy loam      & 272.1            \\ \hline
North Gower clayey loam & 221.9            \\ \hline
Graneville loam         & 651.1            \\ \hline
\end{tabular}
\caption{Equivalent stiffness for common terrain types. A cross-country vehicle is considered with $W=4905$ $N$, $D=0.777$ $m$ and $b = 0.265$ $m$}
\label{terrains}
\end{table}

The motion equations for the general QC can be defined by the following set of differential equations
\begin{equation}
\begin{cases}
m_s\ddot{y}_1+c\left(\dot{y}_1-\dot{y}_2\right)+k\left(y_1-y_2\right)=0 \\
m_{ns}\ddot{y}_2+c\left(\dot{y}_2-\dot{y}_1\right)+k\left(y_2-y_1\right)+k_{tot}\left(y_2-h\right)=0
\end{cases}
\label{EqDiff}
\end{equation}
where $h$ is the terrain elevation profile (please refer to Fig. \ref{AugmentedQC}) and
\begin{equation}
k_{tot}(k_s)=\frac{k_sk_t}{k_s+k_t}
\label{ktot}
\end{equation}
is the combined stiffness that considers the series connection of $k_t$ and $k_s$. From equation (\ref{EqDiff}), one can obtain the state transition model
\begin{equation}
\mathbf{\dot{x}}(t)=f(\mathbf{x}(t),\mathbf{u}(t),t)
\label{DynSys}
\end{equation}
$\mathbf{x}(t)$ being the state vector, $\mathbf{u}(t)$ the input to the system and $f(.)$ the state evolution function. In this research, the following state vector is assumed (omitting $t$)
\begin{equation}
\mathbf{x}=
\begin{pmatrix}
y_1-y_2 \\ \dot{y}_1 \\y_2-h \\ \dot{y}_2 \\ k_{tot}
\end{pmatrix}
\label{StateVector}
\end{equation}
It should be noted that, following this choice for the state vector, both state variables and time-varying parameters (i.e., terrain stiffness) can be concurrently estimated and updated over time during operations in contrast to standard observers where the model parameters are assumed ``static", instead. Joint estimation fits well for estimation problems where the state vector dimension is not large (e.g. in this case). If an estimation of more parameters was required, dual estimation approaches might be faster.\\
The proposed method aims to estimate terrain stiffness through acceleration measurements provided by the onboard sensor suite. Two accelerometers are required that measure the sprung and unsprung mass accelerations, $\ddot{y}_1$ and $\ddot{y}_2$, respectively. The following measurement vector $\mathbf{z}$ can be introduced
\begin{equation}
\mathbf{z}=
\begin{pmatrix}
\ddot{y}_1 \\ \ddot{y}_2
\end{pmatrix}
\label{RispVector}
\end{equation}
and a measurement equation can be drawn in compact matrix form as
\begin{equation}
\mathbf{z}(t)=h(\mathbf{x}(t),t)
\label{RispSys}
\end{equation}
Equations (\ref{DynSys}) and (\ref{RispSys}) form the basis of a non-linear estimation problem that will be solved using the CKF.

\section{Estimation problem}

States that govern dynamic systems can not be always measured directly due to cost and/or technical limitations. Time variation of some parameters of the vehicle model, external disturbances and approximations in the mathematical model may lead to uncertainties. In addition, sensor measurements may be corrupted by noise and biases. Therefore, a stochastic closed-loop observer is required. For its performance and computational efficiency, Kalman Filtering represents a feasible solution.
\begin{figure}[hbtp]
\centering
\includegraphics[scale=0.45]{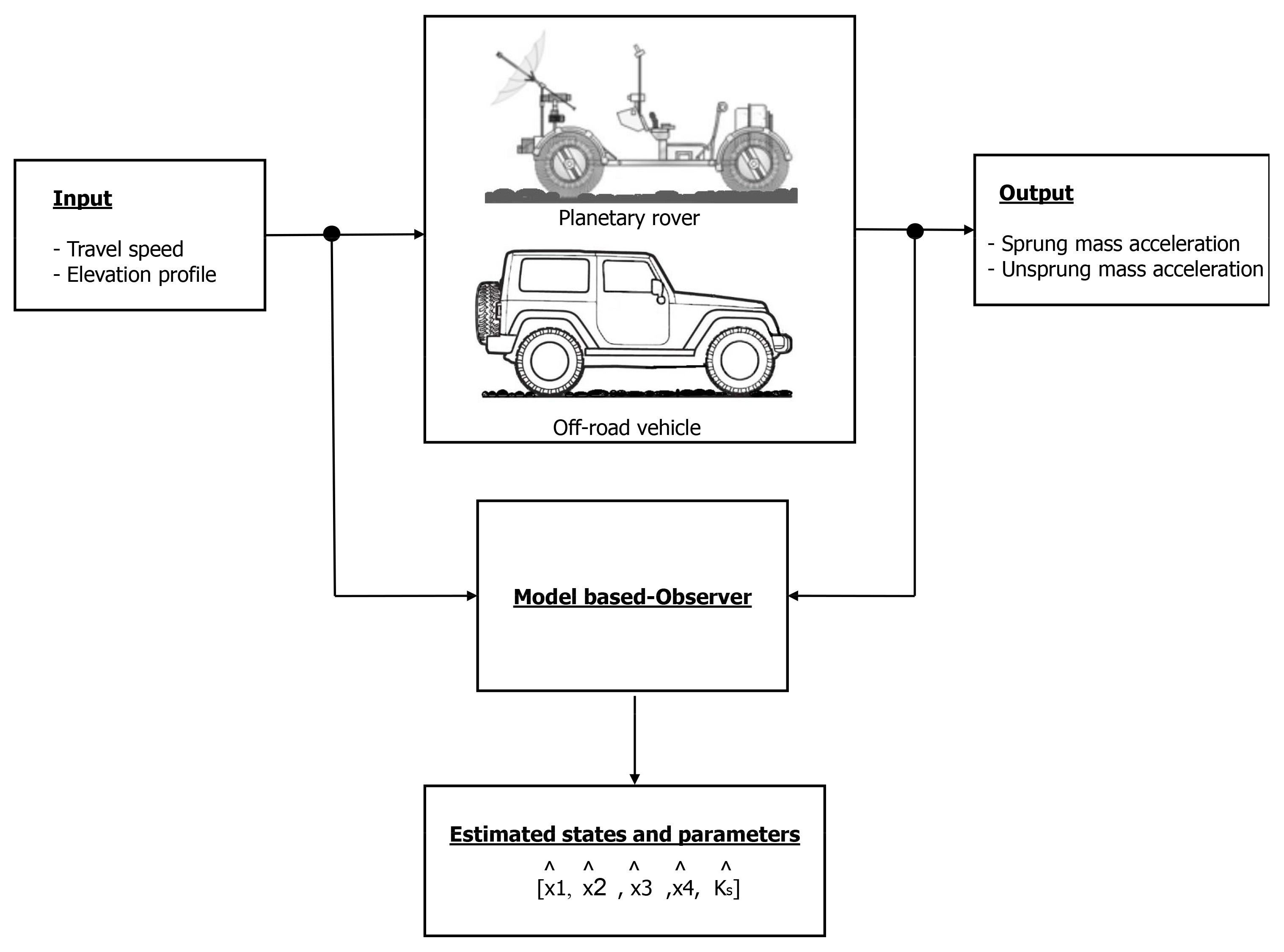}
\caption{Block  diagram of the proposed approach for state/parameter estimation  running in parallel to the system}
\label{Block}
\end{figure}
Figure \ref{Block} shows the block diagram of the proposed framework, where the Kalman filter-based observer runs in parallel with the system. The travel speed, the terrain elevation profile and measurements of the vehicle response in terms of acceleration of the sprung and unsprung mass are fed into the estimator that recursively estimates the motion states (e.g., suspension travel) and parameters of interest (e.g., $k_{s}$) online during normal driving. It should be noted that the proposed approach assumes the availability of a sensing strategy to measure the degree of terrain irregularity in front of the vehicle. This task can be be achieved using exteroceptive sensing with a Lidar \citep{FER} or stereovision \citep{STEREO1}.

A model-based observer can be developed using equations (\ref{DynSys}) and (\ref{RispSys}) that are corrupted by process noise $\mathbf{w}(t)$ and measurement noise $\mathbf{v}(t)$, respectively,
\begin{align}
&\mathbf{\dot{x}}=\mathbf{A}(\mathbf{x})\mathbf{x}+\mathbf{B}\mathbf{u}+\mathbf{G}\mathbf{w}  \label{ourSys2}  \\
&\mathbf{z}=\mathbf{C}(\mathbf{x})\mathbf{x}+\mathbf{v} \label{AccSys2}
\end{align}
where: $\mathbf{A}(\mathbf{x})$ is the transition matrix, $\mathbf{B}$ is the input matrix, $\mathbf{C}(\mathbf{x})$ is the measurement gain matrix and $\mathbf{G}$ is the matrix associated with the process noise. These matrices are developed in the Appendix.
Note that process and sensor noise are assumed to be zero-mean white Gaussian distributions, temporally independent of each other, with auto-covariance matrix respectively $\mathbf{Q}=\text{diag}[\sigma_1^2,\sigma_2^2,\sigma_3^2,\sigma_4^2,\sigma_5^2]$ and $\mathbf{R}=\text{diag}[\sigma_a^2,\sigma_a^2]$. In this work, the following values are assumed
\begin{align*}
&\mathbf{Q}=\text{diag}[10^{-5}m^2,10^{-3}(m/s)^2,10^{-5}m^2,10^{-3}(m/s)^2,10^5(N/m)^2]\\
&\mathbf{R}=\text{diag}[0.5 (m/s^2)^2, 0.5 (m/s^2)^2]
\end{align*}
where process noise parameters are chosen heuristically based on the physical understanding of the problem by the authors, whereas $\mathbf{R}$ elements are usually provided by the sensor manufacturer.



The remainder of this section includes two subsections: the first part is devoted to observability and discretization of the  system, whereas the second part refers to the \textit{Cubature Kalman Filter} algorithm.

\subsection{Observability and discretization}

\subsubsection*{Observability}
A Kalman filter built around a system with unobservable states can not work, because no information may be obtained through the observation vector $\mathbf{z}$ and, consequently, the estimation does not converge on any meaningful solution. Being equation (\ref{ourSys2}) nonlinear, the observability matrix $\mathbf{M}_{\text{obs}}$ is built via Lie derivatives of the observations $\mathbf{z}(t)$ expressed in the form of equation (\ref{RispSys}). By denoting with $n$ the dimension of the state vector, the \textit{i-th} row of the observability matrix $\mathbf{M}_{\text{obs}}^{(i)}$ is built as
\begin{equation}
\forall i \in \mathbb{N}|0\leq i\leq n-1:\mathbf{M}_{\text{obs}}^{(i)}= \frac{\partial}{\partial x}L_f^{(i)}[h(x)]
\end{equation}
where
\begin{align*}
&L_f^{(0)}[h(x)]=h(x) \\
&L_f^{(1)}[h(x)]=\frac{\partial h}{\partial x}f(x,u) \\
&L_f^{(i)}[h(x)]=\frac{\partial}{\partial x}\left[ L_f^{(i-1)}[h(x)]\right] f(x,u)
\end{align*}
The system is observable if and only if the observability matrix has full rank (alias $n$). \\
By performing this analysis on the system under investigation, it results \textit{observable}.

\subsubsection*{Discretization}
It is necessary to express the nonlinear model in a stochastic discrete-time state-space representation.
By solving the differential equation (\ref{ourSys2}) in the sample interval $\Delta t=t_k-t_{k-1}$, one obtains the following discrete system
\begin{align}
&\mathbf{x}_{k+1}=\mathbf{A}_k(\mathbf{x}_k)\mathbf{x}_k+\mathbf{B}_k\mathbf{u}_k+\mathbf{G}_k\mathbf{w}_k \label{DiscSys}\\
&\mathbf{z}_k=\mathbf{C}_k(\mathbf{x}_k)\mathbf{x}_k+\mathbf{v}_k \label{DiscRisp}
\end{align}
with
\footnotesize
\begin{equation}
\mathbf{A}_k=e^{\mathbf{A}\Delta t}, \hspace*{0.3cm}
\mathbf{B}_k=\mathbf{B}\int_{0}^{\Delta t}{e^{\mathbf{A}(\Delta t-\tau)}d\tau}, \hspace*{0.3cm} \\
\mathbf{G}_k=\mathbf{G}\int_{0}^{\Delta t}{e^{\mathbf{A}(\Delta t-\tau)}d\tau}, \hspace*{0.3cm}
\mathbf{C}_k=\mathbf{C}(k,\mathbf{x}_k)
\label{ExactSol}
\end{equation}
\normalsize
where $\mathbf{A}_k$ is obtained by Taylor expansion of $e^{\mathbf{A}\Delta t}$, and $\mathbf{B}_k$ and $\mathbf{G}_k$ are derived by integrating $\mathbf{A}_k$ and multiplying respectively for $\mathbf{B}$ and $\mathbf{G}$. It is possible to achieve a very high accurate discretization with limited computational efforts by stopping the series at the fourth order.

The others matrices that should be discretized in the Kalman filter are: the process uncertainty matrix $\mathbf{Q}$ and the measurement uncertainty matrix $\mathbf{R}$. In this case, $\mathbf{R}_k=\mathbf{R}$, whereas $\mathbf{Q}_k$ may be approximated as suggested in \cite{POW}: $\mathbf{Q}_k\simeq\Delta t\mathbf{G}_k\mathbf{Q}\mathbf{G}_k^T$.

%

\subsection{Cubature Kalman filter}

As an alternative to the standard EKF, the CKF was shown to offer some advantages: better stability and estimation capacity in presence of strong nonlinearities and, furthermore, it does not require to compute any Jacobian matrix. Rather than the pure CKF, the Square-root Cubature Kalman Filter (SCKF) variant is adopted in this research that ensures computational stability when the estimated covariance matrices are not positive definite. The SCKF exploits \textbf{qr} decomposition to obtain the square-root factor $\mathbf{S}_{k|k}$ of the error covariance matrix $\mathbf{P}_{k|k}$, necessary to evaluate cubature points.
The prediction-correction cycle of the SCKF algorithm is reported below, for $i\in \mathbb{N}|1\leq i\leq 2n$, with $n$ the dimension of the state vector.

\textit{\textbf{prediction}} \\
Evaluate the cubature points:
\begin{equation}
\mathbf{X}_{i,k|k}=\mathbf{S}_{k|k}\xi_i +\hat{\mathbf{x}}_{k|k}
\end{equation}
Evaluate the propagated cubature points:
\begin{equation}
\mathbf{X}_{i,k+1|k}^{*}=\mathbf{A}_k(\mathbf{X}_{i,k|k})\mathbf{X}_{i,k|k}+\mathbf{B}_k\mathbf{u}_k
\end{equation}
Estimate the predicted state:
\begin{equation}
\hat{\mathbf{x}}_{k+1|k}=\frac{1}{2n}\sum_{i=1}^{2n}{\mathbf{X}_{i,k+1|k}^{*}}
\end{equation}
Build the weighted centered matrix:
\begin{equation}
\mathbf{\chi}_{k+1|k}^{*}=\frac{1}{\sqrt{2n}}\left[\mathbf{X}_{1,k+1|k}^{*}-\hat{\mathbf{x}}_{k+1|k} \hspace*{0.35cm}...\hspace*{0.35cm} \mathbf{X}_{2n,k+1|k}^{*}-\hat{\mathbf{x}}_{k+1|k}\right]
\end{equation}
Estimate the square-root factor of the predicted error covariance:
\begin{align}
&\mathbf{\zeta}_{k+1|k}=\left[\mathbf{\chi}_{k+1|k}^{*}\hspace*{0.35cm}\sqrt{\mathbf{Q}_k}\right] \notag \\
&\left[\mathbf{q}_{k+1|k}\hspace*{0.35cm}\mathbf{r}_{k+1|k}\right]=\text{qr}\left(\mathbf{\zeta}_{k+1|k}^T\right) \\
&\mathbf{S}_{k+1|k}=\mathbf{r}_{k+1|k}^T \notag
\end{align}

\textit{\textbf{correction}} \\
Update the cubature points:
\begin{equation}
\mathbf{X}_{i,k+1|k}=\mathbf{S}_{k+1|k}\xi_i +\hat{\mathbf{x}}_{k+1|k}
\end{equation}
Update the propagated cubature points:
\begin{equation}
\mathbf{Z}_{i,k+1|k}=\mathbf{C}_k(\mathbf{X}_{i,k+1|k})\mathbf{X}_{i,k+1|k}
\end{equation}
Update the predicted measurement:
\begin{equation}
\hat{\mathbf{z}}_{k+1|k}=\frac{1}{2n}\sum_{i=1}^{2n}{\mathbf{Z}_{i,k+1|k}}
\end{equation}
Build the weighted centered matrix:
\begin{equation}
\mathbf{\mathcal{Z}}_{k+1|k}=\frac{1}{\sqrt{2n}}\left[\mathbf{Z}_{1,k+1|k}-\hat{\mathbf{z}}_{k+1|k} \hspace*{0.35cm}...\hspace*{0.35cm} \mathbf{Z}_{2n,k+1|k}-\hat{\mathbf{z}}_{k+1|k}\right]
\end{equation}
Estimate the square-root factor of the innovation covariance:
\begin{align}
&\mathbf{\zeta}_z=\left[\mathbf{\mathcal{Z}}_{k+1|k}\hspace*{0.35cm}\sqrt{\mathbf{R}_{k+1}}\right] \notag \\
&\left[\mathbf{q}_{z,k+1|k}\hspace*{0.35cm}\mathbf{r}_{z,k+1|k}\right]=\text{qr}\left(\mathbf{\zeta}_{z,k+1|k}^T\right) \\
&{S}_{z,k+1|k}=\mathbf{r}_{z,k+1|k}^T \notag
\end{align}
Build the weighted centered matrix:
\begin{equation}
\mathbf{\chi}_{k+1|k}=\frac{1}{\sqrt{2n}}\left[\mathbf{X}_{1,k+1|k}-\hat{\mathbf{x}}_{k+1|k} \hspace*{0.35cm}...\hspace*{0.35cm} \mathbf{X}_{2n,k+1|k}-\hat{\mathbf{x}}_{k+1|k}\right]
\end{equation}
Update the cross-covariance matrix:
\begin{equation}
\mathbf{P}_{xz,k+1|k}=\mathbf{\chi}_{k+1|k}\mathbf{\mathcal{Z}}_{k+1|k}^T
\end{equation}
Update the Kalman filter gain:
\begin{equation}
\mathbf{W}_{k+1}=\left(\mathbf{P}_{xz,k+1|k}/\mathbf{S}_{z,k+1|k}^T\right)/\mathbf{S}_{z,k+1|k}
\end{equation}
Estimate the updated state:
\begin{equation}
\hat{\mathbf{x}}_{k+1|k+1}=\hat{\mathbf{x}}_{k+1|k}+\mathbf{W}_{k+1}\left(\mathbf{z}_{k+1}-\hat{\mathbf{z}}_{k+1|k}\right)
\end{equation}
Estimate the square-root factor of the updated error covariance:
\begin{align}
&\mathbf{\zeta}_{k+1|k+1}=\left[\mathbf{\chi}_{k+1|k}-\mathbf{W}_{k+1}\mathbf{\mathcal{Z}}_{k+1|k}\hspace*{0.35cm}\mathbf{W}_{k+1}\sqrt{\mathbf{R}_{k+1}}\right] \notag  \\
&\left[\mathbf{q}_{k+1|k+1}\hspace*{0.35cm}\mathbf{r}_{k+1|k+1}\right]=\text{qr}\left(\mathbf{\zeta}_{k+1|k+1}^T\right) \\
&\mathbf{S}_{k+1|k+1}=\mathbf{r}_{k+1|k+1}^T \notag
\end{align}
where $\xi_i=\sqrt{n}\left[\mathbf{I} \hspace*{0.35cm} -\mathbf{I}\right]$ with $\mathbf{I}\in \mathbb{R}^{n\text{x}n}$ identity matrix,
$\mathbf{q}$ orthogonal matrices and $\mathbf{r}$ upper triangular matrices coming from $\mathbf{\zeta}$ matrices; therefore, $\mathbf{S}$ are lower triangular matrices and $\mathbf{P}=\mathbf{S}\mathbf{S}^T$.

\section{Results}

In this section, results are presented to evaluate the performance of the proposed method for combined motion state and soil stiffness estimation. In the first part, the SCKF is compared against the standard EKF in terms of accuracy and stability for different operating conditions. For the reader's convenience, the EKF algorithm is recalled in the Appendix. More details can be found in the Literature, for example in \cite{REI4}. Then, a sensitivity analysis of the SCKF is presented to assess the ability to differentiate between various terrains and the impact of excitation conditions. A reference off-road heavy vehicle is considered, whose parameters are listed in Table \ref{Parameters}. In the second part of this section, the proposed approach is applied to the case of lightweight and low speed lunar roving vehicle.

In the simulations, the true states are obtained by discretizing (time-step of 0.01 s) Equation (\ref{DynSys}), using the correct value of terrain stiffness. Then, process and measurement noise were added to the system in order to feed the estimator with realistic acceleration readings.
\begin{table}[hbtp]
\centering
\begin{tabular}{|l|l|}
\hline
Parameter                             & Value                                 \\ \hline
Suspension stiffness                   & $k=25 \hspace*{0.15cm} kN/m$                    \\ \hline
Suspension damping                     & $c=2 \hspace*{0.15cm} kNs/m$                    \\ \hline
Sprung mass                            & $m_s= 455 \hspace*{0.15cm} kg$                          \\ \hline
Unsprung mass                          & $m_{ns}=45.5 \hspace*{0.15cm} kg$                       \\ \hline
Tire stiffness                         & $k_t=175 \hspace*{0.15cm} kN/m$                  \\ \hline
\end{tabular}
\caption{Parameters of the quarter-car model for a typical off-road vehicle. Please refer to Fig. \ref{AugmentedQC} for more details.}
\label{Parameters}
\end{table}

\subsection{Performance comparison of the SCKF with the EKF}
\begin{figure}[]
\centering
\subfigure[]{\includegraphics[scale=0.27]{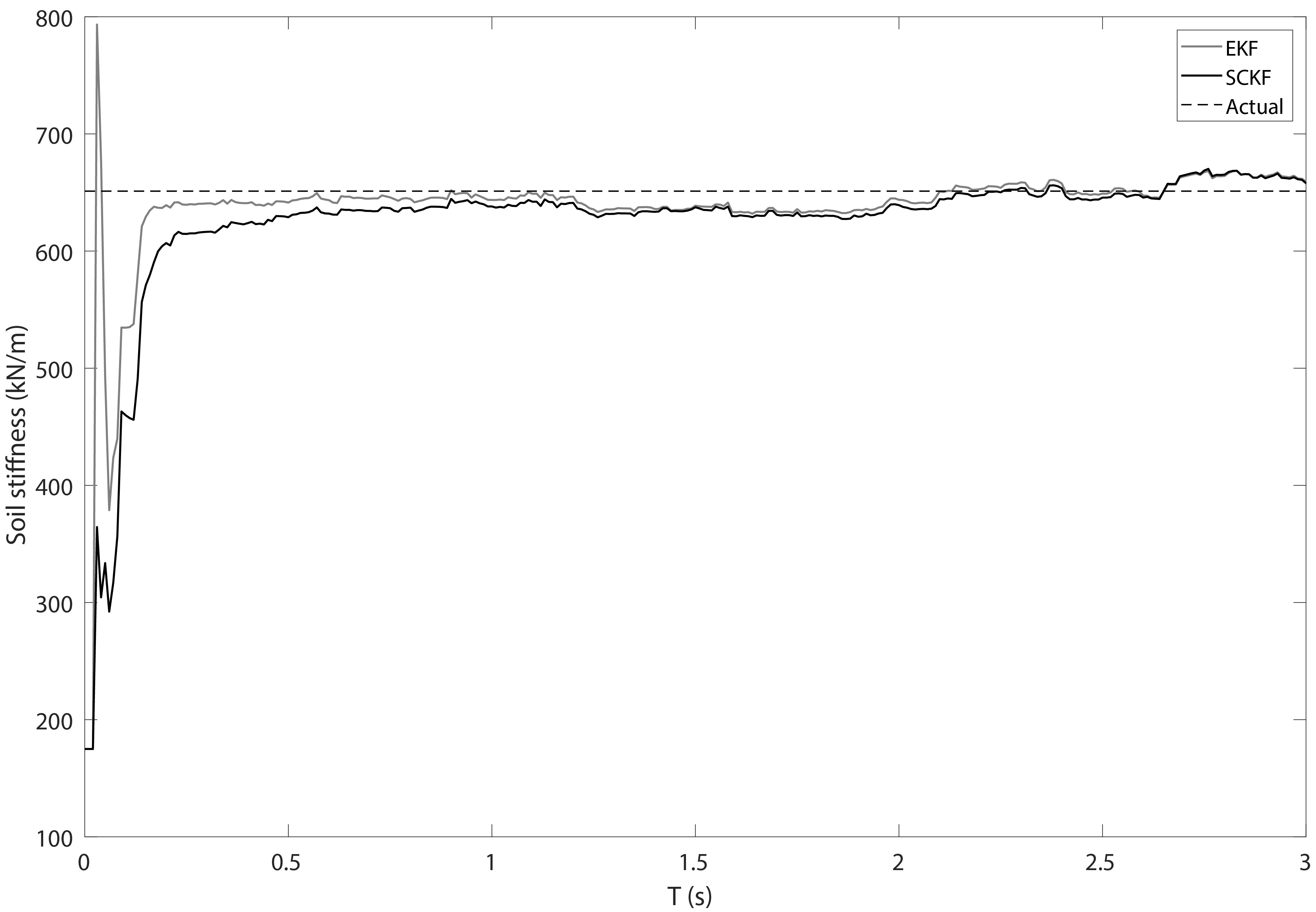}}
\subfigure[]{\includegraphics[scale=0.27]{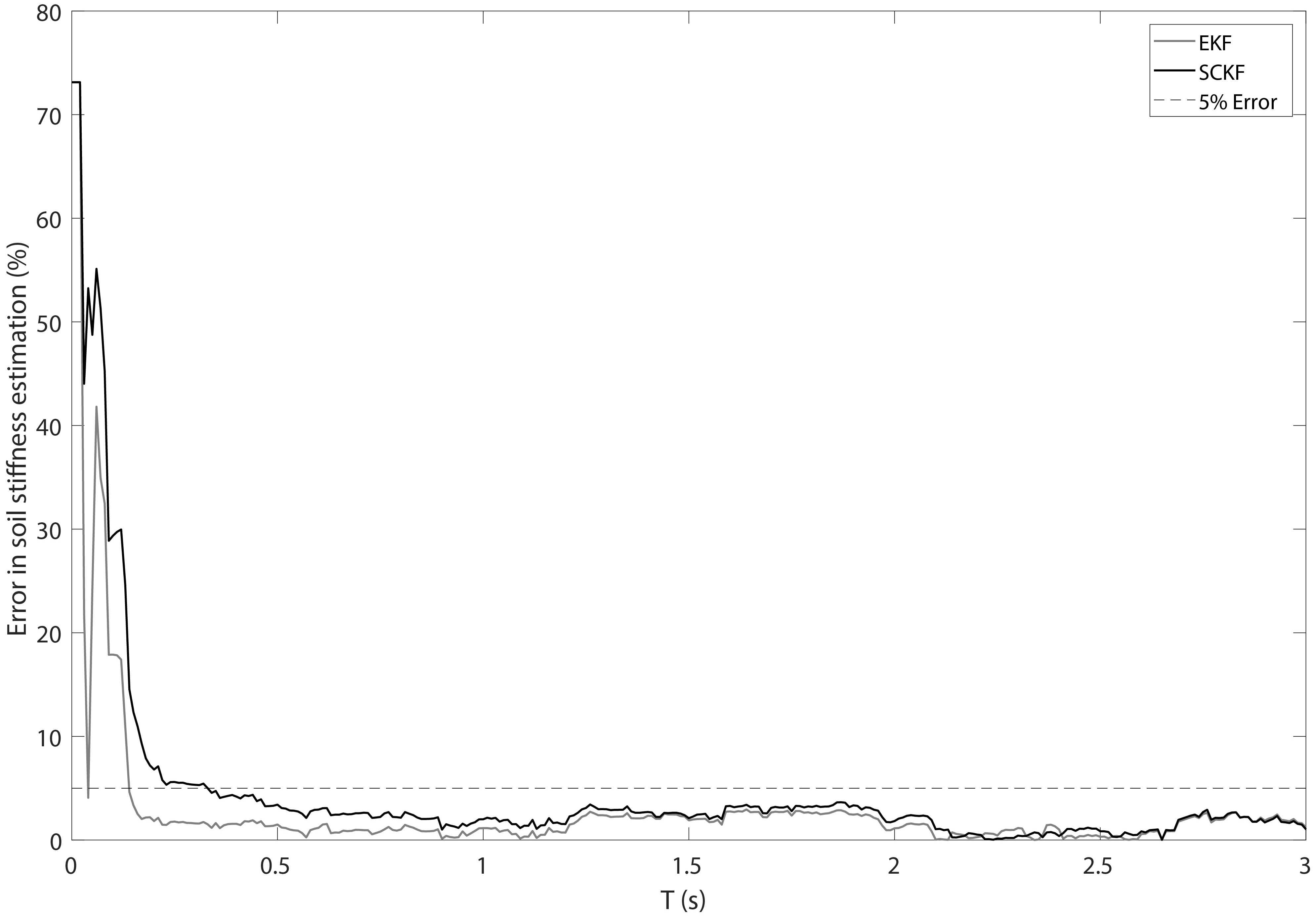}}
\caption{(a) Terrain stiffness estimation as obtained from the EKF and SCKF observers for \textit{Graneville loam} with \textit{ISO-D} profile and 10 m/s vehicle travel velocity.(b) Corresponding estimation error.}
\label{EstimTimeDiff}
\end{figure}

\begin{figure}[]
\centering
\includegraphics[scale=0.3]{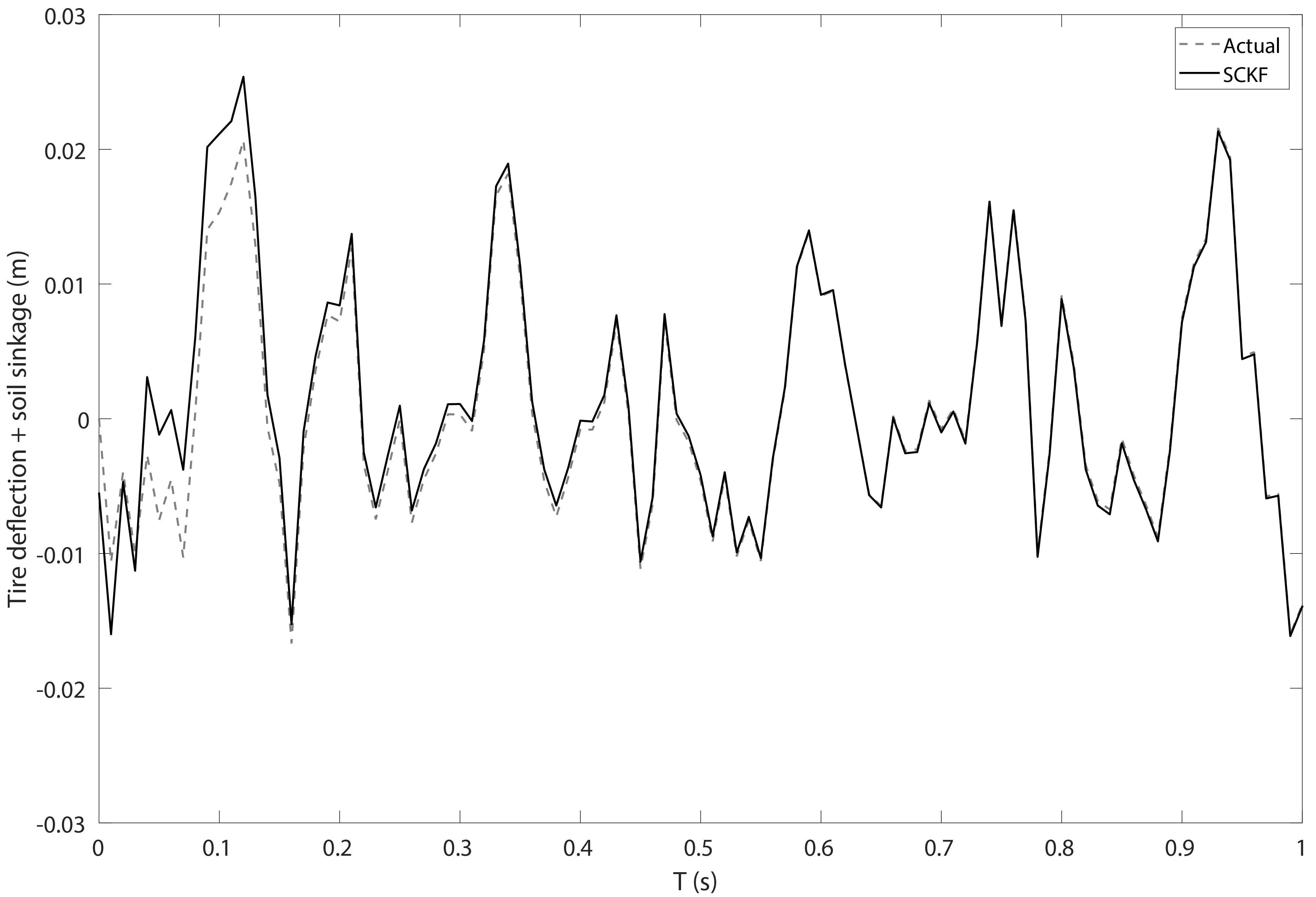}
\caption{State estimation as obtained from the SCKF in the first second of observation. Only tire deflection is shown for simplicity's sake.}
\label{States}
\end{figure}

The SCKF filter and the EKF counterpart are used to estimate soil stiffness as the vehicle travels at constant speed of 10 m/s on a medium hard terrain (i.e., \textit{Graneville loam}) with a low-medium elevation profile \citep{ISO8608} (i.e., \textit{ISO D}). They do not have any a priori information about the current soil stiffness. Therefore, a starting guess equal to half of the tire stiffness is assumed for both algorithms. Figure \ref{EstimTimeDiff}(a) shows soil stiffness estimation as obtained by the EKF observer denoted by a solid grey line compared to that obtained by the SCKF estimator that is marked by a  solid black line. The two observers quickly correct the soil stiffness to the actual value reducing the estimation error to less than 5\% in about 0.5 s (Figure \ref{EstimTimeDiff}(b)). At the same time, estimation of the state variables (solid black line) becomes more and more accurate as shown in Figure \ref{States}, and the discrepancy with respect to the true value (grey dashed line) decreases. For simplicity's sake in Figure \ref{States}, only one of the four states that is the tyre deflection $\mathbf{x}(3)$ is shown as estimated from the SCKF during the first second of the simulation. The EKF provides very similar results and they are omitted here.

\begin{figure}[]
\centering
\includegraphics[scale=0.3]{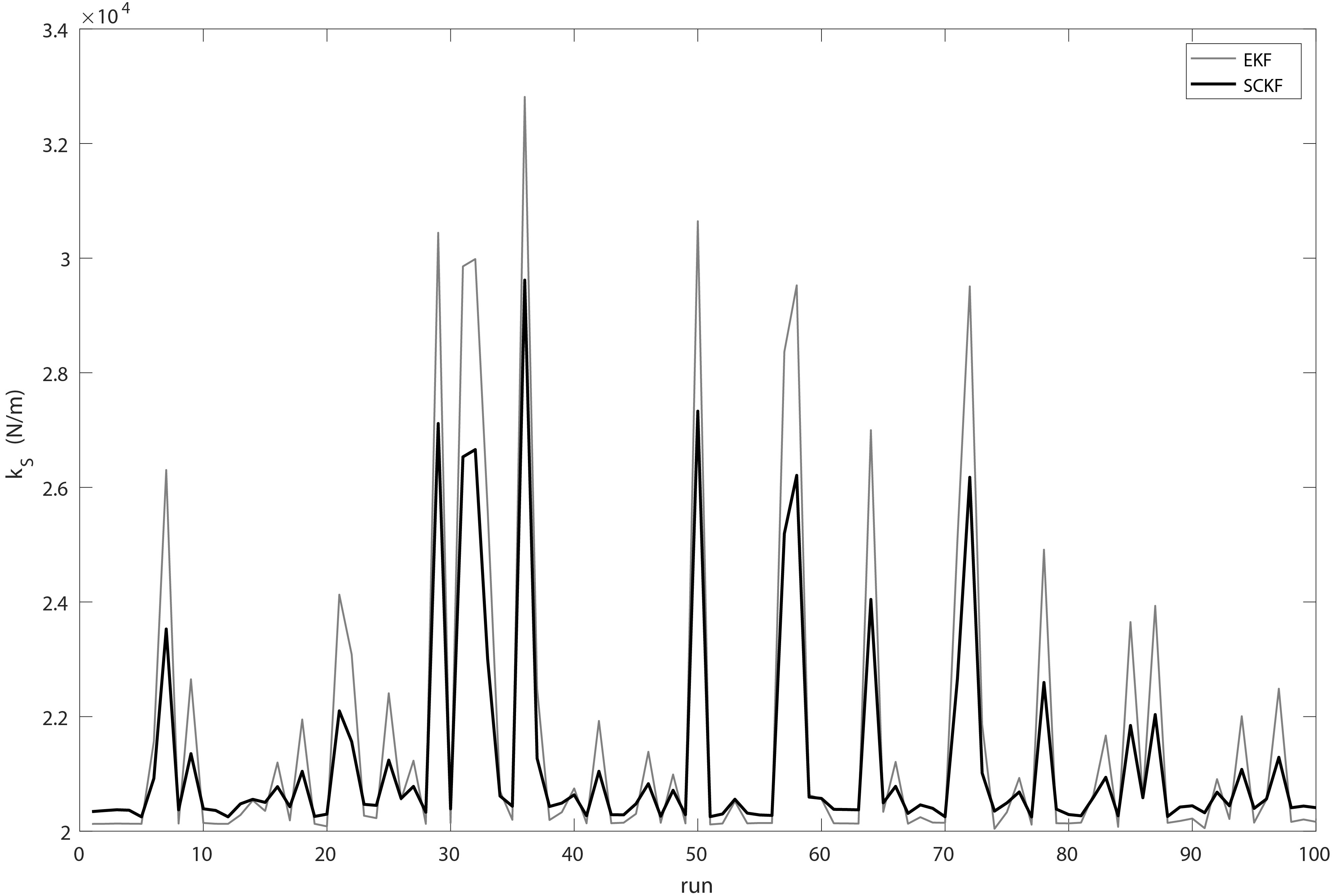}
\caption{Comparison between SCKF and EKF in terms of RMSE in the stimation of $k_{s}$ on \textit{Graneville loam}, elevation profile \textit{ISO-D} and $10 \hspace*{0.1 cm} m/s$  vehicle velocity.}
\label{AEKFvsASCKF}
\end{figure}

\begin{figure}[hbtp]
\centering
\includegraphics[scale=0.35]{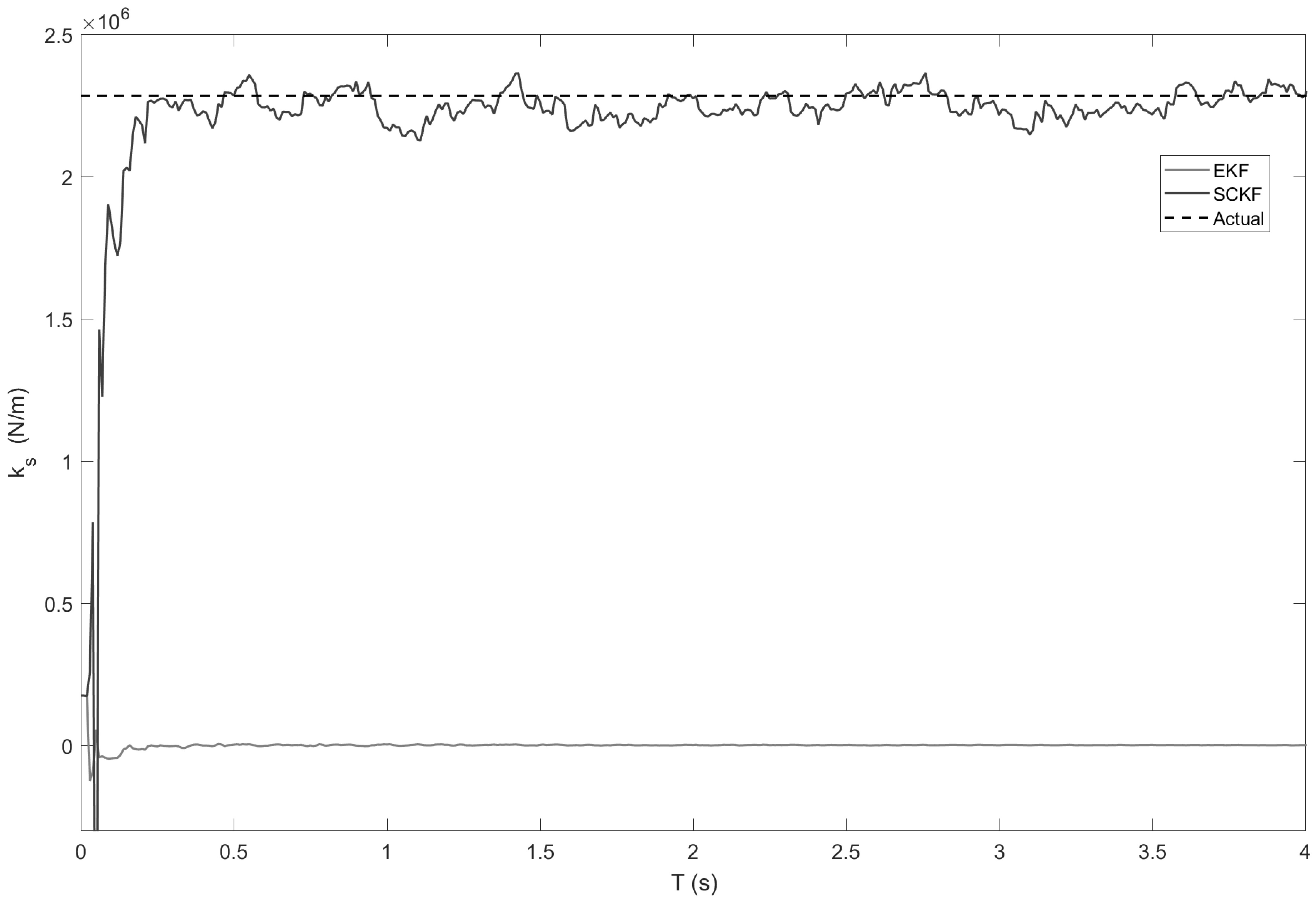}
\caption{Terrain stiffness estimation as obtained from the EKF and SCKF observers for \textit{LETE sand} with \textit{ISO-F} elevation profile and 10 m/s vehicle travel velocity.}
\label{StabilityDiff}
\end{figure}

As seen from Figure \ref{EstimTimeDiff}, the performance of the SCKF are comparable with those of the standard EKF filter.
In order to get a quantitative evaluation in terms of estimation accuracy, the root-mean square error (RMSE) is evaluated considering an inspection window of 10 s for 100 Monte Carlo simulations. Results are collected in Figure \ref{AEKFvsASCKF}. The SCKF (solid black line) slightly outperforms the EKF in terms of accuracy with a relative average error of 3.26\% and 0.28\% standard deviation against 3.33\% and 0.45\% standard deviation.\\
For the given terrain properties and travel speed, the EKF showed no instability episode.
However, under harsher conditions, that is, hard \textit{LETE sand} with an \textit{ISO - F} elevation profile and equal travel speed of $10 $ m/s, the EKF observer fails to converge to the correct value of terrain stiffness, as shown in Figure \ref{StabilityDiff}. In contrast, the SCKF algorithm correctly estimates the terrain property.  Similar instabilities are observed when considering other operating conditions, thus, confirming that the SCKF preserves stability on a wider range of excitations than the EKF counterpart.

Summing up, Table \ref{Comparison} collects the strengths and weaknesses of the two observers. It results that the SCKF outperforms the EKF in two out of three metrics, with similar performance in the third case. Therefore, the SCKF is chosen as the preferred filter in this research, whereas the EKF is not retained further.

\begin{table}[hbtp]
\centering
\begin{tabular}{|c|c|c|}
\hline
\textbf{Performance}  & \textbf{EKF} & \textbf{SCKF} \\ \hline
Velocity   & \checkmark     & \checkmark      \\ \hline
Accuracy              & \textbf{}     & \checkmark     \\ \hline
Stability             & \textbf{}     & \checkmark      \\ \hline
\end{tabular}
\caption{Comparative table between the EKF and SCKF observers.}
\label{Comparison}
\end{table}

\subsection{Sensitivity analysis}

First, the ability of the SCKF to differentiate between various surfaces is evaluated. Then, the impact of the operating conditions on the observer performance is studied.

\begin{figure}[]
\centering
\includegraphics[scale=0.35]{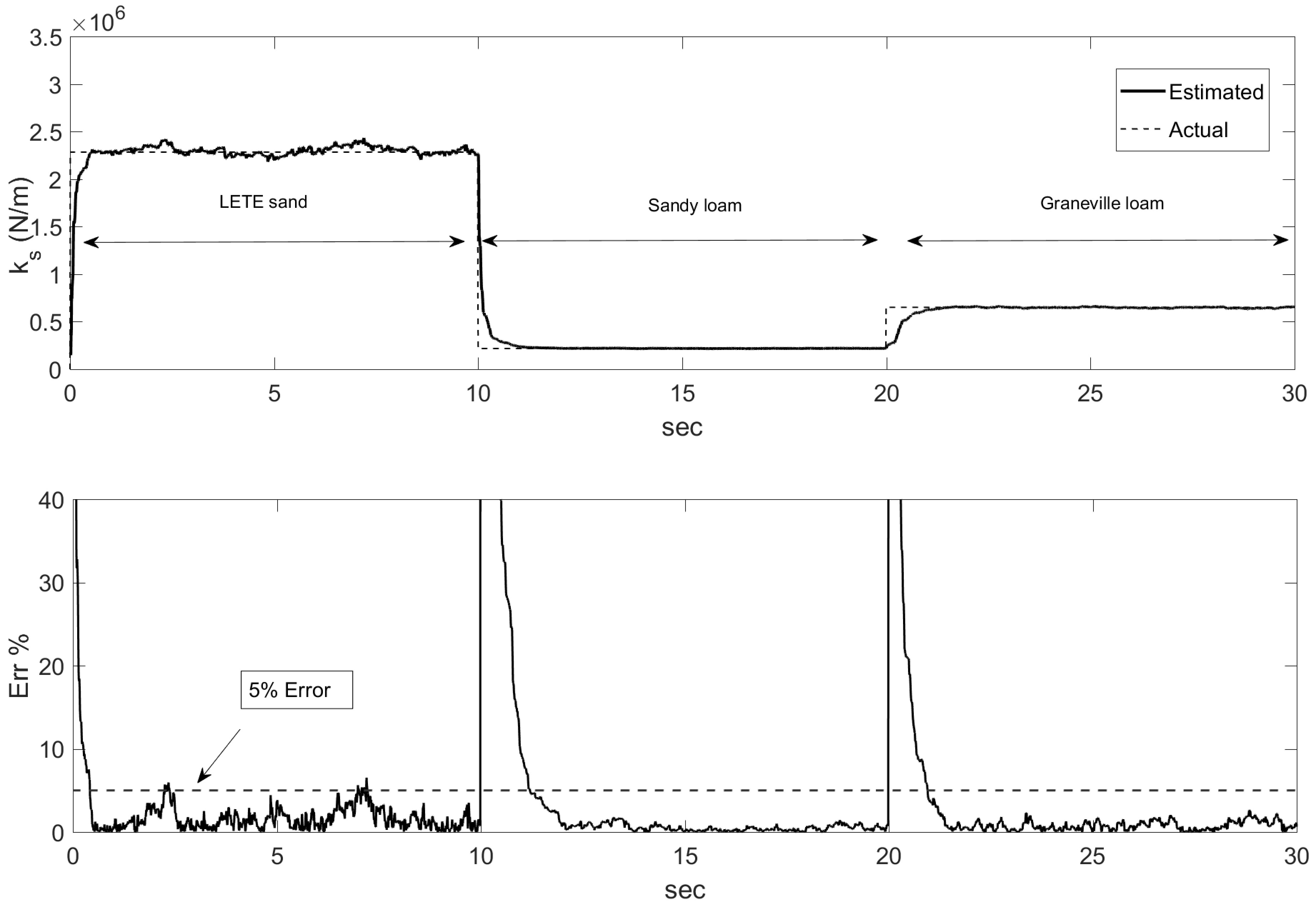}
\caption{Terrain stiffness estimation as obtained from the SCKF observer along a mixed-terrain path: \textit{LETE sand}, \textit{Upland sandy loam}, \textit{Graneville loam}. An off-road elevation profile \textit{Meadow-c} and a velocity of $10 \hspace*{0.1cm} m/s$  are assumed.}
\label{Exp1}
\end{figure}

When travelling in natural environments, the supporting surface can vary substantially ranging from gravel, to loam or sand, etc. As a consequence, soil stiffness is subject to temporal changes especially for long range and long duration applications that include precision agriculture and planetary exploration.
A representative path is built by assembling in sequence stretches of different uniform surfaces, namely \textit{LETE sand}, \textit{Upland sandy loam} and \textit{Graneville loam}. As the elevation profile, a harsh \textit{Meadow c} \citep{MEADOW} is assumed for all terrains, whereas the travel speed is set equal to 10 m/s. Soil stiffness estimation as obtained from the SCKF is shown in the upper part of Figure \ref{Exp1}. The observer correctly detects the actual terrain properties (dashed black line) with an adaptation window of a few seconds during both the two surface transitions: the first from hard \textit{LETE sand} to ``soft"  \textit{Upland sandy loam}, and the second where two surfaces with similar characteristics are considered (i.e., \textit{Upland sandy loam} and \textit{Graneville loam}). The estimation accuracy is good with a relative error that mostly falls well under 5\% (bottom plot of Figure \ref{Exp1}).

A second simulation set is now presented to assess the impact of the system excitation, i.e. terrain elevation profile and vehicle travel speed, keeping unchanged the previous terrain sequence as input.

\begin{figure}[hbtp]
\centering
\includegraphics[scale=0.4]{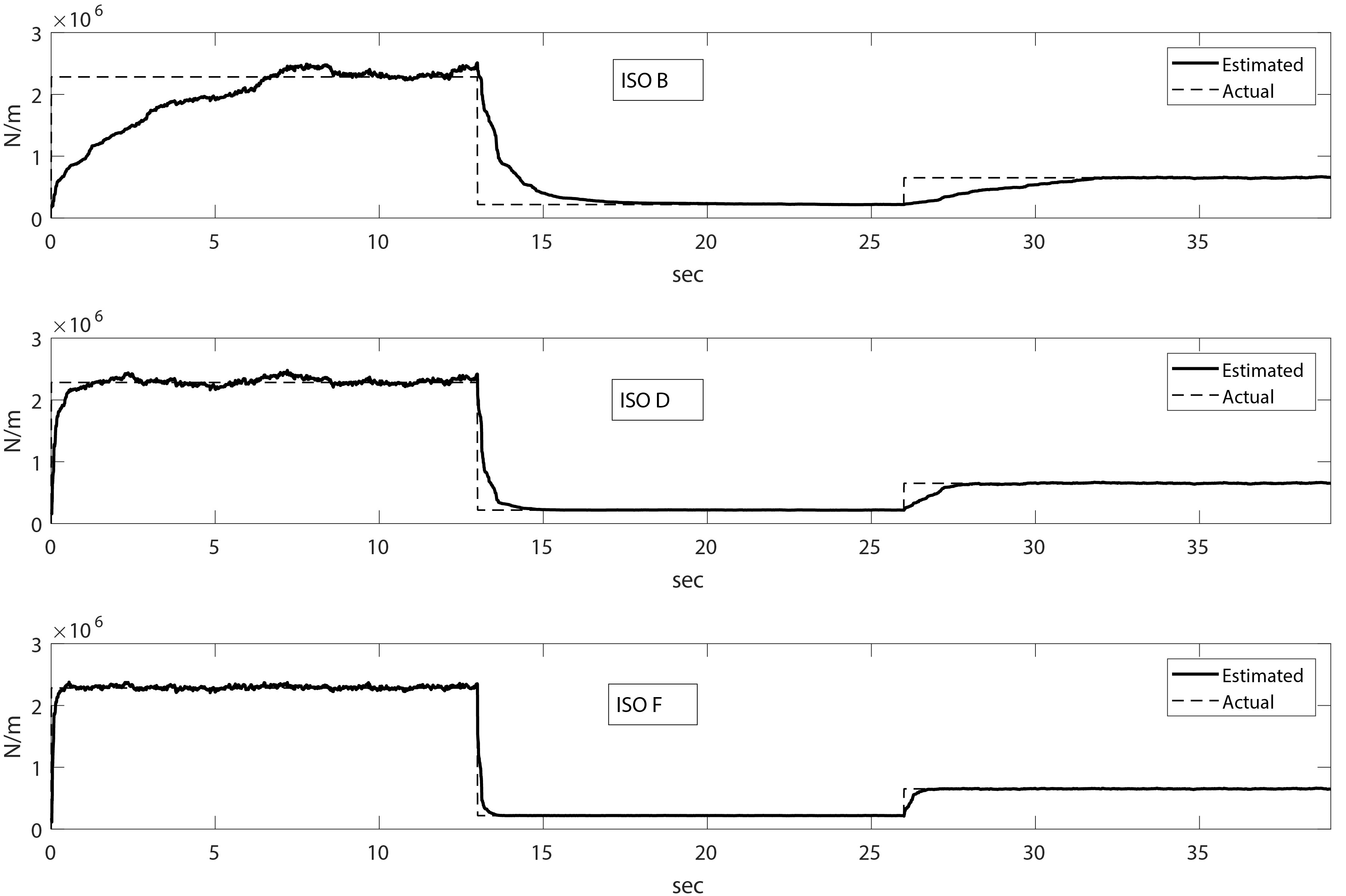}
\caption{$k_s$ estimation on a path with three different terrains and a given travel velocity, by varying \textit{ISO} profile among: \textit{B} (upper plot), \textit{D} (center plot) and \textit{F} (bottom plot)}
\label{Exp2}
\end{figure}

\begin{figure}[hbtp]
\centering
\includegraphics[scale=0.4]{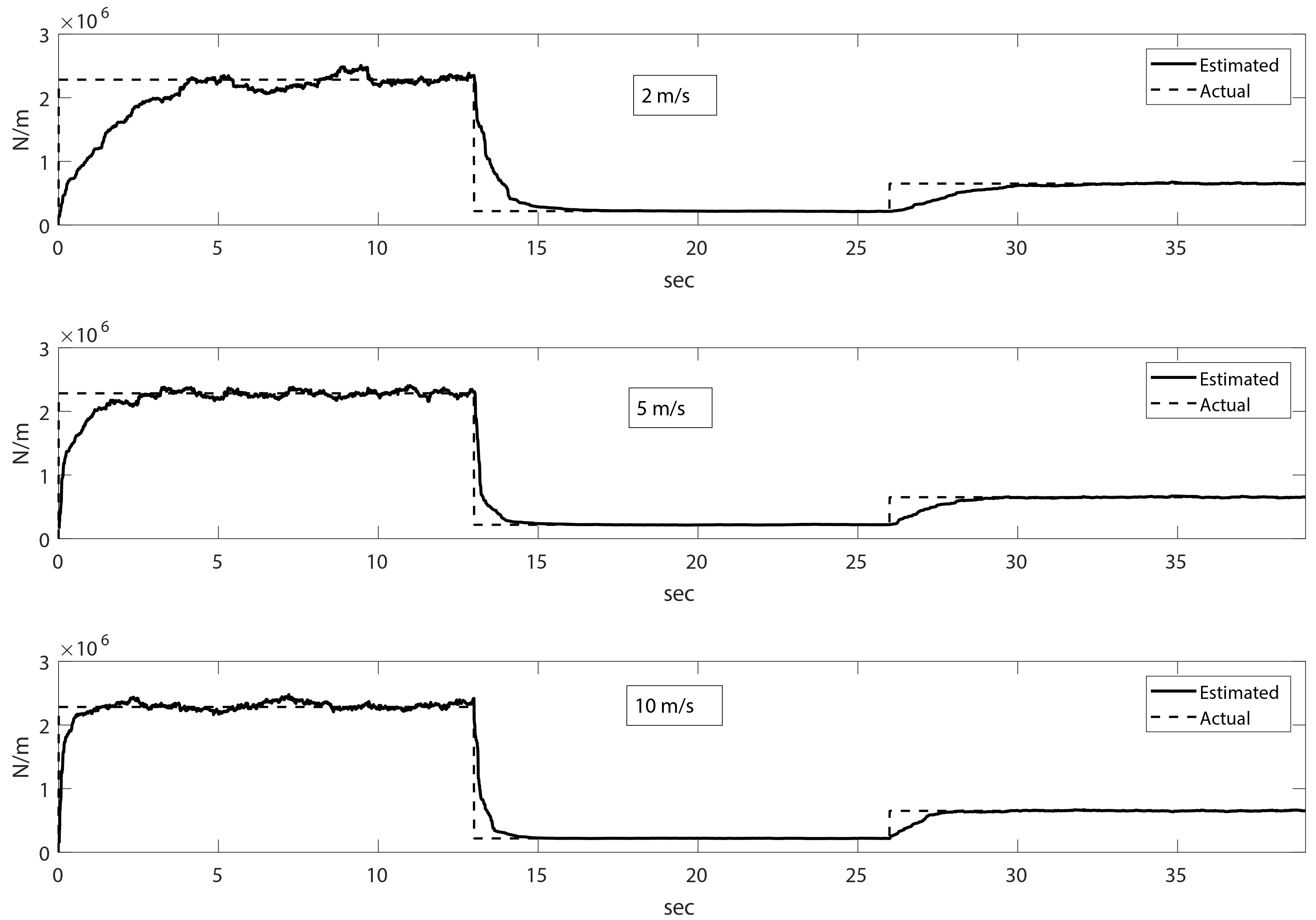}
\caption{$k_s$ estimation on a path with three different terrains and a given soil profile, by varying vehicle velocity among: $2$ (upper plot), $5$ (center plot) and $10$ (bottom plot) $m/s$}
\label{Exp3}
\end{figure}

In Figure \ref{Exp2} three elevation profiles with increasing roughness degree (i.e., \textit{ISO B}, \textit{D} and \textit{F}) are considered at a constant vehicle velocity (i.e., $10 \hspace*{0.1cm} m/s$). The smoother the profile, the less accurate is the estimation. The relative percentage RMSE decrases from 9.18\% to 1.06\% when moving from ISO B to ISO F. Similarly, the average adaptation window reduces from 5.54 $s$ to 0.51$s$.\\
Figure \ref{Exp3} shows the estimation results when the vehicle velocity increases (from $2\hspace*{0.1cm} m/s$, to $5\hspace*{0.1cm} m/s$, and to $10\hspace*{0.1cm} m/s$), as the elevation profile is kept constant (\textit{ISO D}). The larger the travel velocity, the faster and more accurate the estimation, with a relative percentage RMSE moving from 5.83\% for $2\hspace*{0.1cm}m/s$ to 2.28\% for $10\hspace*{0.1cm}m/s$ vehicle velocity.\\
For an extensive evaluation of the system performance under different excitation conditions, one can refer to the average percentage error as of Table \ref{errperc}, considering just the traverse of \textit{Graneville loam}. As seen from this table, terrain roughness plays a more important role than vehicle velocity.

\begin{table}[hbtp]
\centering
\begin{tabular}{|c|c|c|c|}
\hline
\multicolumn{4}{|c|}{\textbf{Average percentage error}}       \\ \hline
\multicolumn{2}{|c|}{$v = 10 m/s$} & \multicolumn{2}{c|}{ISO D} \\ \hline
Profile         & Value          & Velocity      & Value      \\ \hline
ISO B           & 9.18\%         & $2\hspace*{0.15cm}m/s$         & 5.83\%     \\ \hline
ISO D           & 2.28\%         & $5\hspace*{0.15cm}m/s$         & 2.60\%     \\ \hline
ISO F           & 1.06\%         & $10\hspace*{0.15cm}m/s$        & 2.28\%     \\ \hline
\end{tabular}
\caption{Average percentage error under different excitation conditions for \textit{Graneville loam}.}
\label{errperc}
\end{table}

\begin{figure}[]
\centering
\includegraphics[scale=0.2]{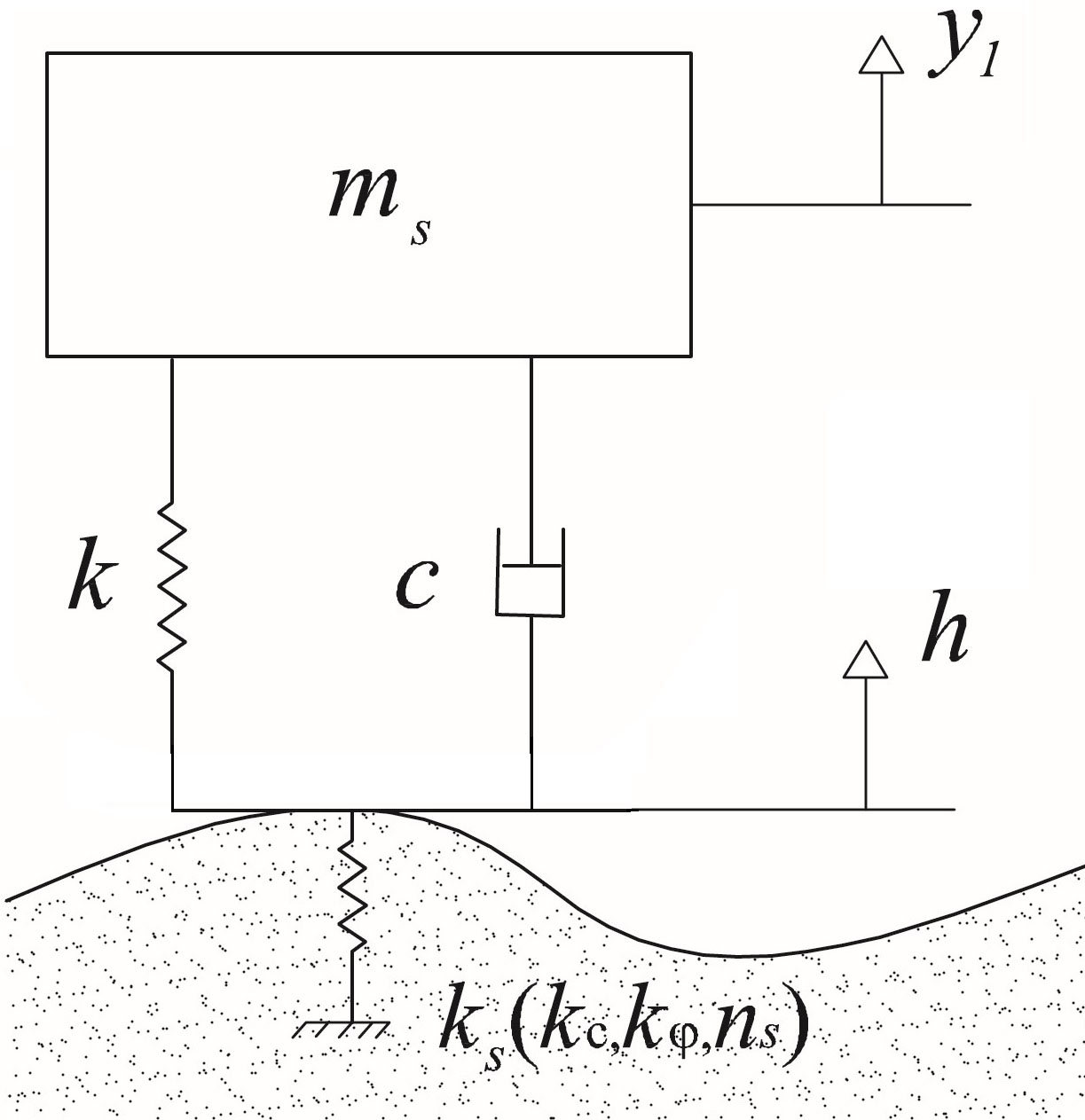}
\caption{General 1 DOF QC model for the Lunar rover vehicle}
\label{Fig10}
\end{figure}

\subsection{Application to planetary roving vehicles}
Finally, the proposed method is applied to the ``Moon buggy" that is the Lunar Roving Vehicle (LRV) used in the Apollo program. The LRV is a relatively lightweight and low-speed vehicle that was employed by astronauts for moving and equipment transport on lunar soil. In this vehicle, no coil spring is present and the suspension and tire stiffness may be combined into a joined stiffness $k$, as well as, sprung and unsprung mass can be condensed in a unique mass $m_s$, as shown in Figure \ref{Fig10}. This leads to a single degree of freedom system, as suggested in \cite{LRV1}, whose model parameters are reported in Table \ref{LRVQC}.\\ The state and measurement vector of the corresponding estimation problem reduce respectively to $\mathbf{x}=[y_1-h\hspace*{0.25cm} \dot{y_1}\hspace*{0.25cm} k_{tot}]^T$ and $\mathbf{z}=[\ddot{y_1}]$, whereas the transition matrix $\mathbf{A}$ can be obtained from that of the 2 DOF QC case by removing the third and fourth line and column, and matrices $\mathbf{B}$, $\mathbf{C}$, $\mathbf{G}$ can be adapted accordingly. Corresponding process and measurement noise are left unchanged.\\
The SCKF-based observer is implemented onboard the LRV to evaluate the equivalent stiffness of the lunar soil \textit{Regolith} that is equal to $k_s$=28.487 kN/m, considering the LRV geometry ($D=1.016$ $m$ and $b=0.254$ $m$ \citep{WheelLRV}) and vertical load on Earth ($W=735$ $N$). In this simulation, a constant vehicle velocity equal to $2\hspace*{0.1cm}m/s$ is considered (close to the lower bound of the velocity range). Although a statistical model of the lunar terrain irregularity according to the ISO standards has not been found in the literature, a reasonable assumption is that the terrain can be considered ``very poor" (ISO Class G).
\begin{table}[hbtp]
\centering
\begin{tabular}{|l|l|}
\hline
Parameter        & Value             \\ \hline
Stiffness        & $k=15\hspace*{0.15cm}kN/m$      \\ \hline
Damping          & $c=1.5\hspace*{0.15cm}kNs/m$      \\ \hline
Mass             & $m_s=75\hspace*{0.15cm}kg$        \\ \hline
Velocity range   & $v= 1-5\hspace*{0.15cm}m/s$        \\ \hline
\end{tabular}
\caption{Parameters of the single DOF quarter-car model for the LRV.}
\label{LRVQC}
\end{table}

\begin{figure}[]
\centering
\includegraphics[scale=0.35]{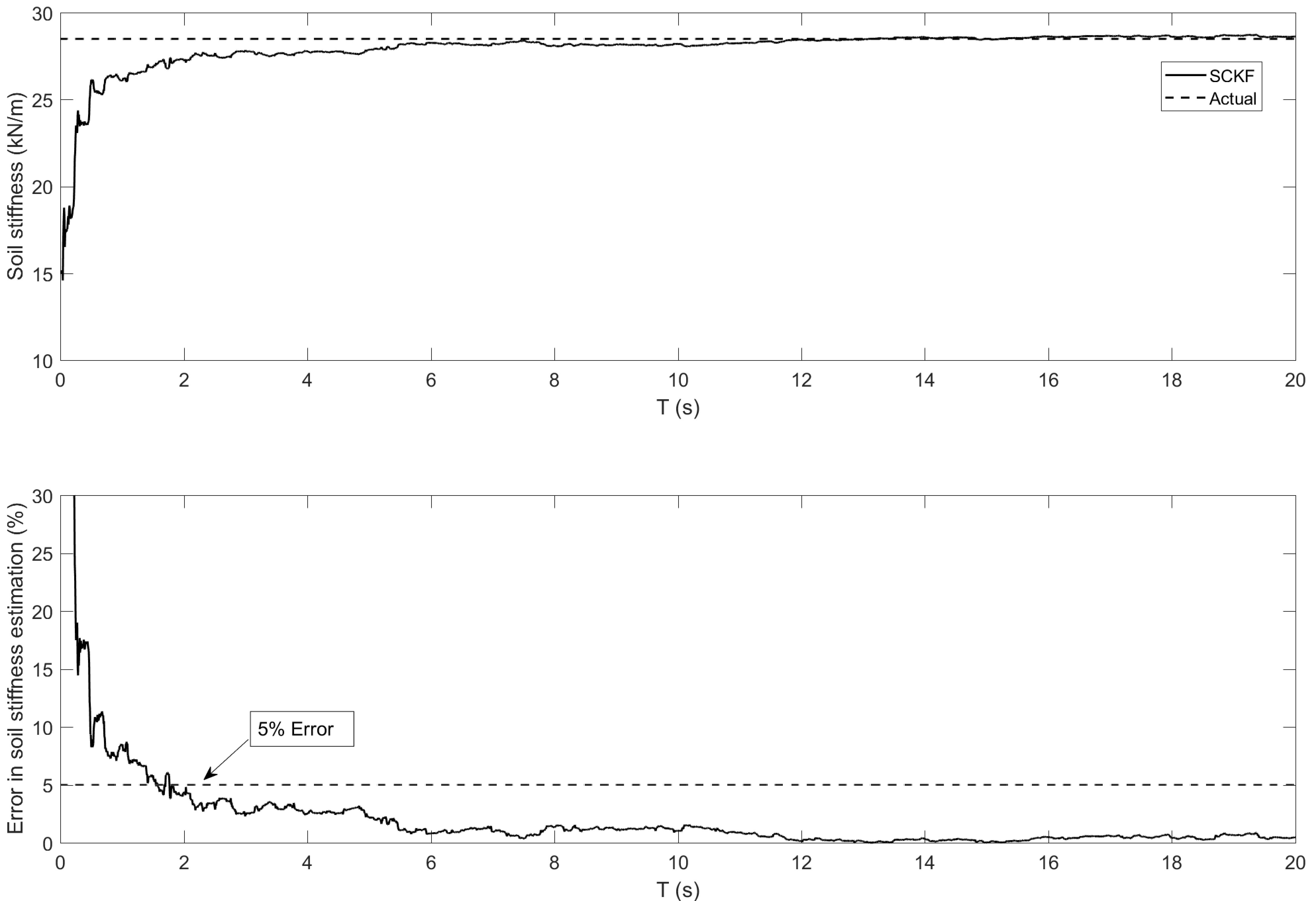}
\caption{(Upper plot) Terrain stiffness estimation as obtained from the SCKF observer for regolith with ISO-G profile and 2 m/s travel velocity of the Moon buggy. (Lower plot) Corresponding estimation error.}
\label{Fig11}
\end{figure}

Results are shown in Figure \ref{Fig11}. Although the scenario under investigation is substantially different in terms of vehicle and terrain properties and travel speed, the SCKF confirms its ability in estimating the equivalent regolith stiffness with an error less than 5\% after about 2 s of observation and less than 1\% after about 13 s. This figure verifies the general applicability of the proposed observer to a wide family of rough-terrain vehicles ranging from heavy to lightweight, and medium to low speed.

\section{Conclusions}

In this paper, a nonlinear model-based observer was presented grounded in the cubature integration rule to estimate terrain equivalent stiffness in rough terrain vehicles. The SCKF was compared with the de-facto standard EFK showing slightly better performance in terms of accuracy and a higher stability to varying operation conditions. Its implementation does not require the computation of Jacobian matrices that can be rather complex especially for highly nonlinear systems, thus ensuring computationally efficiency.\\
An extensive sets of simulations were presented to quantitatively evaluate the filter performance for varying travel speed, terrain irregularity and deformability. Even under low excitation conditions (e.g., low speed on very smooth surfaces), the SCKF estimated terrain properties with an error always below 5\%. The applicability to a wide range of vehicles was also investigated showing that the proposed approach performs well for both heavy and lightweight and fast and slow vehicles.\\
This research opens the way to different developments: firstly, more complex models may be considered in order to include pitch and roll motions, e.g, half car and full car. Secondly, as the SCKF approach allows strong nonlinearities to be studied in a more efficient way than classical nonlinear Kalman filter, alternative terrain model could be considered that do not expressly consider  the linearization of the terrain constitutive equation.

\section{Appendix}

\appendix
\setcounter{equation}{0}
\renewcommand{\theequation}{A.\arabic{equation}}

In this Appendix, the matrices presented in Equations (\ref{ourSys2}) and (\ref{AccSys2}) are developed in detail and the EKF algorithm is recalled.

\subsection{Model matrices and noise vector development}

\begin{equation}
\mathbf{A}(\mathbf{x})
=\begin{bmatrix}
0 & 1 & 0 & -1 & 0 \\
-\frac{k}{m_s} & -\frac{c}{m_s} & 0 & \frac{c}{m_s} & 0 \\
0 & 0 & 0 & 1 & 0 \\
\frac{k}{m_{ns}} & \frac{c}{m_{ns}} & -\frac{\mathbf{x}(5)}{m_{ns}} & -\frac{c}{m_{ns}} & 0 \\
0 & 0 & 0 & 0 & 0
\end{bmatrix}
\end{equation}
\begin{equation}
\mathbf{B}
=\begin{bmatrix}
0 \\ 0 \\ -1 \\ 0 \\ 0
\end{bmatrix}
\end{equation}
\begin{equation}
\mathbf{G}
=\begin{bmatrix}
0 & 1 & 0 & 1 & 0 \\
1 & 1 & 0 & 1 & 0 \\
0 & 0 & 0 & 1 & 0 \\
1 & 1 & 1 & 1 & 0 \\
0 & 0 & 0 & 0 & 1
\end{bmatrix}
\end{equation}
\begin{equation}
\mathbf{C}(\mathbf{x})
=\begin{bmatrix}
-\frac{k}{m_s} & -\frac{c}{m_s} & 0 & \frac{c}{m_s} & 0 \\
\frac{k}{m_{ns}} & \frac{c}{m_{ns}} & -\frac{\mathbf{x}(5)}{m_{ns}} & -\frac{c}{m_{ns}} & 0 \\
\end{bmatrix}
\end{equation}
\begin{equation}
\mathbf{w}
=\begin{pmatrix}
w_1 \\ w_2 \\ w_3 \\ w_4 \\ w_5
\end{pmatrix}
\end{equation}
\begin{equation}
\mathbf{v}
=\begin{pmatrix}
v_a \\ v_a
\end{pmatrix}
\end{equation}
with: $\forall i \in \mathbb{N}|1\leq i \leq 5:w_i\sim\mathcal{N}(0,\sigma_i^2)$, $v_a\sim\mathcal{N}(0,\sigma_a^2)$, white and independent.

\subsection*{EKF algorithm}

The EKF operates according to the following cycle:
\begin{align}
&\textbf{\textit{prediction}} \notag \\
&\hat{\mathbf{x}}_{k+1|k}=\mathbf{A}_k\hat{\mathbf{x}}_{k|k}+\mathbf{B}_k\mathbf{u}_k \\
&\mathbf{P}_{k+1|k}=\mathbf{A}_{x,k}\mathbf{P}_{k|k}\mathbf{A}_{x,k}^T+\mathbf{Q}_k \notag
\end{align}
\begin{align}
&\textbf{\textit{correction}} \notag \\
&\mathbf{W}_{k+1}=\mathbf{P}_{k+1|k}\mathbf{C}_{x,k+1}^T\left(\mathbf{C}_{x,k+1}\mathbf{P}_{k+1|k}\mathbf{C}_{x,k+1}^T+\mathbf{R}_{k+1}\right)^{-1} \notag \\
&\hat{\mathbf{x}}_{k+1|k+1}=\hat{\mathbf{x}}_{k+1|k}+\mathbf{W}_{k+1}\left(\mathbf{z}_{k+1}-\mathbf{C}_{k+1}\hat{\mathbf{x}}_{k+1|k}\right) \\
&\mathbf{P}_{k+1|k+1}=\left(\mathbf{I}-\mathbf{W}_{k+1}\mathbf{C}_{x,k+1}\right)\mathbf{P}_{k+1|k} \notag
\end{align}
where: $\mathbf{A}_k$, $\mathbf{B}_k$ and $\mathbf{C}_k$ are the discretized process, input and measurement matrices, $\mathbf{I}\in \mathbb{R}^{n\text{x}n}$ the identity matrix, $\mathbf{W}_{k+1}$ the Kalman filter gain, $\hat{\mathbf{x}}_{k+1|k}$ the predicted state vector, $\mathbf{P}_{k+1|k}$ the covariance matrix for $\hat{\mathbf{x}}_{k+1|k}$, $\hat{\mathbf{x}}_{k+1|k+1}$  the updated state vector, and $\mathbf{P}_{k+1|k+1}$  the updated estimate covariance matrix. In these equations, $\mathbf{A}_{x,k}$ is the discretization of the process Jacobian $\mathbf{A}_x(\mathbf{x})$, reported below, and $\mathbf{C}_{x,k+1}$ is the discretization of the measurement Jacobian $\mathbf{C}_x(\mathbf{x})$ given by second and fourth rows of $\mathbf{A}_x(\mathbf{x})$ matrix.
\begin{equation}
\mathbf{A}_x(\mathbf{x})=\frac{\partial}{\partial \mathbf{x}}[\mathbf{A}(\mathbf{x})\mathbf{x}]=
\begin{bmatrix}
0 & 1 & 0 & -1 & 0 \\
-\frac{k}{m_s} & -\frac{c}{m_s} & 0 & \frac{c}{m_s} & 0 \\
0 & 0 & 0 & 1 & 0 \\
\frac{k}{m_{ns}} & \frac{c}{m_{ns}} & -\frac{\mathbf{x}(5)}{m_{ns}} & -\frac{c}{m_{ns}} & -\frac{\mathbf{x}(3)}{m_{ns}} \\
0 & 0 & 0 & 0 & 0
\end{bmatrix}
\label{Jacobian}
\end{equation}

\begin{acks}
The financial support of the projects: Autonomous DEcision making in very long traverses (ADE), H2020 (Grant No. 821988) and Agricultural inTeroperabiLity and Analysis System (ATLAS), H2020 (Grant No. 857125), is gratefully acknowledged.
\end{acks}

\bibliographystyle{SageH}
\bibliography{JVC}

%
%
%

\end{document}